\documentclass[sigconf]{acmart}
\AtBeginDocument{%
  }

\acmYear{2025}
\acmDOI{10.48550/arXiv.2504.07475}

\acmConference{Purposeful XR Workshop. Presented at CHI 2025 (arXiv:2504.07475). Yokohama, Japan}

\acmSubmissionID{CHIXR/2025/09}

\begin{document}

%%
%% The "title" command has an optional parameter,
%% allowing the author to define a "short title" to be used in page headers.
\title{Leveraging Agency in Virtual Reality to Enable Situated Learning}

%%
%% The "author" command and its associated commands are used to define
%% the authors and their affiliations.
%% Of note is the shared affiliation of the first two authors, and the
%% "authornote" and "authornotemark" commands
%% used to denote shared contribution to the research.
\author{Eileen McGivney}
\orcid{0000-0002-3416-7488}
\affiliation{%
  \institution{Northeastern University}
  \city{Boston}
  \state{Massachuetts}
  \country{USA}
}
\email{e.mcgivney@northeastern.edu}

%%
%% By default, the full list of authors will be used in the page
%% headers. Often, this list is too long, and will overlap
%% other information printed in the page headers. This command allows
%% the author to define a more concise list
%% of authors' names for this purpose.

%%
%% The abstract is a short summary of the work to be presented in the
%% article.
\begin{abstract}
Learning is an active process that is deeply tied to physical and social contexts. Yet schools traditionally place learners in a passive role and focus on decontextualizing knowledge. Situating learning in more authentic tasks and contexts typically requires taking it outside the classroom via field trips and apprenticeships, but virtual reality (VR) is a promising tool to bring more authentically situated learning experiences into classrooms. In this position paper, I discuss how one of VR’s primary affordances for learning is heightening agency, and how such heightened agency can facilitate more authentically situated learning by allowing learners legitimate participation. \end{abstract}

%%
%% The code below is generated by the tool at http://dl.acm.org/ccs.cfm.
%% Please copy and paste the code instead of the example below.
%%


%%
%% Keywords. The author(s) should pick words that accurately describe
%% the work being presented. Separate the keywords with commas.
%% A "teaser" image appears between the author and affiliation
%% information and the body of the document, and typically spans the
%% page.

%%\received{20 February 2007}
%%\received[revised]{12 March 2009}
%%\received[accepted]{5 June 2009}

%%
%% This command processes the author and affiliation and title
%% information and builds the first part of the formatted document.
\maketitle

\section{Situated Learning and the Failure of Traditional Schooling}
It has long been recognized that cognition is not something entirely “in the head,” but is deeply connected to context, community, and identity \cite{rogoff_everyday_1984}. For the same reason, it has long been argued that learning cannot mean only passively absorbing content knowledge, but must be an active and participatory process of moving from novice to expert \cite{lave_situated_1991}. Situated and situative learning theories explain how learning is sociocultural, embedded in the context in which occurs, with community, and requiring the learner’s whole self \cite{greeno_commentary_2015, lave_situated_1991}. Yet traditional schooling is devoid of the rich contexts and tasks that help learners meaningfully connect to knowledge and engage in the kind of active participation that will be required of them outside the classroom \cite{gee_situated_2004}. The decontextualized schooling model contributes to young people’s disengagement in education as they lose motivation to learn throughout secondary school, and also makes education less effective by failing to prepare learners to transfer what they learn throughout their future lives \cite{vedderweiss_adolescents_2011, day_import_2012}. On the other hand, authentically situated learning opportunities have been shown to address this important challenge, supporting motivation and learning outcomes \cite{kelley_conceptual_2016, lave_situated_1991}.

Typically, better situating learning in authentic tasks and contexts has required taking it out of the classroom via apprenticeships and field trips \cite{dewitt_short_2008, lave_situated_1991}. However, immersive technologies are promising tools to bring situated learning opportunities into classrooms, making these experiences a more frequent part of school-based education and accessible to learners everywhere \cite{dede_immersive_2009}. In recent years, virtual reality (VR) has become more affordable and accessible, and studies illustrate its affordances for bringing field trips into classrooms to learn about complex phenomena such as the impact climate change has on the ocean \cite{markowitz_immersive_2018}. The promise of VR to promote situated learning in the classroom lies in its ability to make learners feel a sense of presence in a different place than the classroom, but also by giving them agency over their learning to take actions that are difficult or impossible in the real world \cite{makransky_cognitive_2021}. While immersion and presence have been the focus of most research on learning with VR \cite{hamilton_immersive_2021, jensen_review_2018}, its novel forms of interactivity and first-person perspective may be more powerful affordances for situating learning.

\section{Agency in Virtual Reality Learning Environments}
Presence and agency have been identified as the two primary affordances of VR for learning, each contributing to learning outcomes by allowing learners to feel transported to far-off places, visualizing phenomena that they cannot see in real life, heightening their engagement and intrinsic motivation to learn, ultimately increasing a variety of learning outcomes \cite{johnson-glenberg_immersive_2018, makransky_cognitive_2021}. Agency is one’s capacity for acting and exerting control, which studies have shown is heightened in interactive VR systems. For example, studies have compared using an interactive learning environment with passively watching the content in VR and found interactivity heightens learners’ sense of agency and enhances some of their learning outcomes \cite{johnson-glenberg_platform_2021, petersen_study_2022}. Immersive learning models and VR designers have equated the ability to interact with the environment and objects as the equivalent of agency. For example, the Cognitive Affective Model of Immersive Learning (CAMIL) posits that the control factors in a VR experience are its technical feature that directly leads to agency \cite{makransky_cognitive_2021}. 

However, my research highlights the need to focus on more than just the ability to move in the environment and interact with objects to give learners the type of agency needed for authentically situated learning opportunities. Situated learning explicitly requires legitimate participation, which goes beyond using your body but requires the ability to take meaningful actions that progressively move you into an expert role \cite{lave_situated_1991}. Similarly, other educational models of agency focus on supporting learners’ autonomy to set goals and make decisions about what and how to learn \cite{ryan_self-determination_2000}, which requires a broader definition of control and interaction than moving objects \cite{mcgivney_interactivity_2025}. In my research, I find that defining agency as interactivity is problematic because they have a complex association. For one, agency has multiple dimensions, and using less-interactive media like 360-degree videos still gives learners agency over what they focus their attention on and feeling in control over their learning despite not being able to move and interact with the environment \cite{mcgivney_interactivity_2025}. Further, giving learners more opportunity to interact in a VR field trip does not necessarily enhance their sense of agency \cite{kruger_complexity_2025}.

\subsection{Transforming Learning through Agentic Experiences in Virtual Reality}

These exploratory studies of agency in relation to varied levels of interactivity suggest more thought needs to be given to designing VR experiences for the types of agency that are beneficial for actively engaging learners in situated learning opportunities. While technological advances are making highly interactive VR experiences more feasible and accessible, it is going to be crucial to embed a learning framework that leverages interactivity in a productive way. My current research is investigating the application of different types of media and experiences within VR such as the balance between guided instruction and unguided discovery in VR, how 360-degree videos can be combined with more interactive activities, and narratives that highlight consequences of their actions over broader storylines. For example, my lab is currently developing a VR field trip that situates physics learning in the context of automotive manufacturing. Learners will not only be able to run car crash simulations and move around the environment, they will also be given control over a day in the life of an automotive engineers, giving them legitimate participation in real-world tasks.

One potential risk of not taking a thoughtful approach to designing for agency is that VR learning experiences will become highly controlled environments. Much research on learning with VR to date has focused on its shortcomings as a medium for didactic instruction, finding that heightened cognitive load makes it more difficult for learners to remember content than if they used a slideshow or video \cite{mcgivney_promoting_2023}. My own research finds that increased sense agency does not lead to increased content learning gains \cite{kruger_complexity_2025}. One response to this body of research would be to make VR a less stimulating environment, including controlling learners’ focus to enhance their content retention. Given advances in biometric data collection and sensing, VR systems could force learners to direct their gaze and attention at specific aspects of the environment. This would make VR a tool well suited to the traditional schooling model that restricts learner agency.

However, this approach ignores the more transformative benefits VR can bring to classrooms. Doing so requires us to focus on learning outcomes beyond content knowledge gains. For example, heightened agency in VR learning environments increases learners’ self-efficacy and intrinsic motivation to learn, and the heightened arousal that may distract learners is also responsible for making them feel a sense of awe and wonder for the material \cite{chirico_effectiveness_2017, mcgivney_promoting_2023, queiroz_efficacy_2023, queiroz_students_2022}. 

Ultimately, VR’s potential to make transformative contributions to education rests on how well it is designed for agency. Bringing more authentically situated learning opportunities to classrooms will help make learning more meaningful, effective, and motivating. But doing so will require focusing on the design of the experiences beyond making them increasingly immersive and interactive, and rather focusing on giving learners opportunities for legitimate participation.

%% The next two lines define the bibliography style to be used, and
%% the bibliography file.
\bibliographystyle{ACM-Reference-Format}
\bibliography{CHI25Workshop}

\end{document}